\documentclass[12pt,tightenlines,eqsecnum,floats,shownopacs,nofootinbib,amsmath,amssymb,aps,prd]{revtex4}

\usepackage{color}
\usepackage{graphicx}
\usepackage{amsmath,amssymb}

\def\be{\nopagebreak[3]\begin{equation}}
\def\ee{\end{equation}}
\def\ba{\nopagebreak[3]\begin{eqnarray}}
\def\ea{\end{eqnarray}}

\begin{document}

\title{Time in Fundamental Physics}
\author{Abhay Ashtekar}
\email{ashtekar@gravity.psu.edu} \affiliation{Institute for
Gravitation and the Cosmos \& Physics Department, Penn State,
University Park, PA 16802, U.S.A.}

\begin{abstract}

The first three sections of this article contain a broad brush
summary of the profound changes in the notion of time in fundamental
physics that were brought about by three revolutions: the
foundations of mechanics distilled by Newton in his
\emph{Principia}, the discovery of special relativity by Einstein
and its reformulation by Minkowski, and, finally, the fusion of
geometry and gravity in Einstein's general relativity. The fourth
section discusses two aspects of yet another deep revision that
waits in the wings as we attempt to unify general relativity with
quantum physics.

\end{abstract}

\maketitle

\section{Newton's abstraction and its success}
\label{s1}

We perceive the passage of time through change. However the
existence of change by itself does not establish the reality of an
\emph{objective} time. Indeed, our direct observations refer to a
\emph{relational} notion of time.
For example, routine measurements only tell us that while the earth
goes around the sun once, the moon goes around the earth
approximately 13 times. Or, while the second hand completes one
round on one's wristwatch, one's pulse beats 70 times. What we
directly experience is change and observations only let us compare
durations involved in one change with those involved in another.
This led to what is often referred to as the Leibnitzian space-time
view in which only meaningful questions about motion refer to
relative motion.

This notion of relational time has a curious similarly with the
barter system people used before the advent of the abstract concept
of money. A sheep was worth $n$ chickens, a chicken was worth $m$
bottles of oil, and so on. People only \emph{compared} values of
objects. Money ---particularly in the form of banknotes--- is an
abstract concept, a mental creation, that simplifies trade. Money is
not essential for survival. Indeed, one cannot eat it nor can one
run a car engine by filling the gas tank with it. It is a notion
that was devised to simplify book-keeping. One can imagine
abolishing money and using just the barter system that only assigns
relative values to pairs of necessary objects. But the notion of
money is extremely powerful: it \emph{streamlines} all commercial
transactions by giving each item an absolute value in place of the
pairwise relational values used in the barter system. It is
difficult to imagine a flourishing trade without money, let alone
the more abstract monetary instruments that are now used (and that
led to the 2008 economic melt-down!).

Through his \emph{Principia}, Newton streamlined time in the same
fashion. He postulated that \emph{absolute time} in itself has a
direct physical meaning, without reference to any physical systems
or phenomena. To distinguish this notion from other subjective or
psychological measures of the passage of time, he represented the
absolute physical time by a 1-dimensional mathematical continuum.%
\footnote{In the contemporary terminology used in the general
relativity literature, what Newton called absolute space provides a
canonical foliation of space-time and each slice is labeled by a
value of the time parameter (taking values in the one-dimensional
affine space $\mathbb{R}$). The preferred family of observers are
the Galilean ones, in uniform motion with respect to one another.
They all agree on this parametrization. What I call Newtonian
space-time in this article is generally referred to as
\emph{neo-Newtonian} or \emph{Galilean} space-time in the history
and philosophy of science circles following \cite{je}. I thank Tom
Pashby for pointing this out.}
All durations were to be measured against this absolute time. Newton
taught us that we need not be satisfied with \emph{relational time},
e.g., just with thinking of how many rotations of the moon around
the earth correspond to one rotation of the earth around the sun.
Rather, the moon's orbit around earth marks an interval on the
\emph{physical} one dimensional continuum of time by itself, and so
does the earth's orbit around the sun. Any inertial observer can
measure these intervals and Newtonian mechanics asserts that they
will all agree in spite of the uniform relative motion. Thus, the
flow of time is absolute in spite of Galilean relativity. \emph{If
we wish}, we can compare the lengths of these intervals and conclude
that there are approximately 13 moon-orbit intervals in one
earth-orbit interval. And we can also compare these intervals with
those marked on the absolute time continuum by the rotation of the
second hand on one's watch or one's pulse. But the comparisons and
the resulting relational notion of time are all secondary. Absolute
time is the primary, physical notion.

This notion lies at the foundation of Newton's laws of mechanics.
The resulting celestial mechanics were astonishingly successful.
Already in the 1750s, papers appeared in the Philosophical
Transactions of the Royal Society, calculating sophisticated
consequences of Newton's laws, such as of the effect of the
gravitational pull of Jupiter and Saturn on earth's motion (see,
e.g.,\cite{ptrs})! Had one continued to use the relational time of
direct experience ---as Leibnitz, for example, advocated--- such
calculations would have become very cumbersome. Celestial mechanics
could be developed and its predictions could be compared against
observations so quickly largely because it was based on Newton's
absolute time.

The success of celestial mechanics brought home the message that the
heavens are not so mysterious after all; they were brought within
the grasp of the human mind. Thus, the Principia shattered
Aristotelian orthodoxy by abolishing the distinction between heaven
and earth. For the first time, there were truly universal
principles. An apple falling on earth and the planets orbiting
around the sun were now subject to the same laws, formulated using
the same absolute time continuum. No wonder then that Principia
literally sculpted human consciousness, providing the mental images
that people commonly use. Most people now think of time as a part of
physical reality, flowing serenely, all by itself, untouched by the
external world.

In spite of this success, Newton was well aware of the fact that
there was no objective basis for his postulates on absolute time and
absolute space. He had to invoke theological arguments in support of
their absolute character. Not surprisingly, Leibnitz criticized
these arguments as untenable.

\section{Special relativity: Abolishing absolute time}
\label{s2}

The Principia quickly replaced Aristotle's four books on physics and
became the new orthodoxy. It reigned supreme for over 150 years.
However, a challenge to the Newtonian world view then emerged from
totally unexpected quarters: advances in the understanding of
electromagnetic phenomena. In the middle of the 19th century, the
Scottish physicist James Clarke Maxwell achieved an astonishing
synthesis of all the accumulated knowledge concerning these
phenomena in just four vectorial equations. These equations further
provided a specific value of the velocity $c$ of light. But this
velocity did not refer to a reference frame; \emph{it appeared as an
absolute constant of Nature}. Now, the notion of an absolute
velocity blatantly contradicts Galilean relativity, a cornerstone on
which Newtonian mechanics rests. This tension between Maxwell's
electrodynamics and Newtonian mechanics dismayed natural
philosophers. But by then learned men had developed deep trust in
the Newtonian world and therefore concluded that Maxwell's equations
can only hold in a specific reference frame, called the
\emph{ether.} The value of the speed of light $c$ that emerged from
Maxwell's equations, they concluded, is relative to this ether. But
by doing so, they in fact reverted back to the Aristotelian view
that Nature specifies an absolute rest frame. A state of confusion
remained for some 50 years.

There were several leading figures such as Henri Poincar\'e and
Hendrik Lorentz who attempted to resolving this tension through
mathematical modifications of the Galilean transformations. However,
it was Albert Einstein who grasped the deep physical implications of
this quandary: \emph{It was asking us to abolish Newton's absolute
time.} In 1905 Einstein accepted the implications of Maxwell's
equations at their face value and used simple but ingenious thought
experiments to argue that, since the speed $c$ of light is a
universal constant, the same for all inertial observers, Newton's
notion of absolute simultaneity is physically untenable. Spatially
separated events which appear as simultaneous to one observer can
not be so for another observer, moving uniformly with respect to the
first. The Newtonian model of space-time can only be an
approximation that holds when speeds involved are all much smaller
than $c$. A new, better model of space-time structure emerged and
with it new kinematics, called \emph{special relativity.}

In special relativity the elementary notion is that of an event
---e.g., the explosion of a firecracker--- which is completely
localized in space and in time. These events constitute the
space-time continuum. Strictly, this is also the case in Newtonian
mechanics. What changes in special relativity is that \emph{time no
longer has a privileged standing.} Already in Newtonian mechanics,
the spatial distance between two \emph{events} separated by time is
not absolute; it depends on the state of motion of the Galilean
observer. In special relativity, time joins space: time intervals
between two distant events also depend on the state of motion of the
observer. Minkowski realized and emphasized the profound implication
of this change: in special relativity, only the \emph{4-dimensional}
space-time continuum and its geometry are observer-independent.%
\footnote{In special relativity, space-time is still represented by
the 4-dimensional affine space $\mathbb{R}^4$. However, there is no
longer a preferred foliation of this continuum into space and time.
Each inertial observer introduces her own foliation and can speak of
space and time intervals between any two events. But as we change
the observer, the values of these intervals change, leaving only the
space-time interval invariant. In the Newtonian space-time of
footnote 1, there is a preferred spatial slice through each event,
labeled by the value of absolute time. In special relativity,
through each event one has instead a light cone, trajectories of
light rays emanating (and converging) at that event. In the limit $c
\to \infty$, these cones go over to the Newtonian spatial slices of
absolute time.}

This new paradigm immediately led to some dramatic predictions.
Energy and mass lost their identity and could be transformed into
one another, subject to the famous equation $E= Mc^2$. Since the
velocity of light $c$ is so large in conventional units, the energy
contained in a gram of matter can therefore illuminate a town for a
year. It predicted that a twin who leaves her brother behind on
earth and goes on a trip in a spaceship traveling at a speed near
the speed of light for a year would return to find that her brother
had aged several decades. The origin of these astonishing
consequences lies in the replacement of Newton's absolute time by
the special relativistic \emph{observer dependent} one. But they are
so counter-intuitive that, as late as the 1930s, there were debates
in prominent western universities whether special relativity could
be philosophically viable. But we know that these misgivings were
all completely misplaced. Nuclear reactors function on earth and
stars shine in the heavens, converting mass into energy, obeying
$E=mc^2$. In high energy laboratories, particles routinely reach
velocities close to that of light and are known to live orders of
magnitude longer than their twins at rest on earth.

More generally, we now routinely encounter billions of applications
of special relativity both in frontier science and in gadgets used
in everyday life. They provide convincing evidence that the absolute
time of Newton's does \emph{not} correspond to physical reality.
It is an approximate concept that is very useful in situations in
which all speeds involved are low compared to that of light (as is
the case in celestial mechanics). Special relativity led us to a
more sophisticated notion in which the absolute distinction between
space and time intervals is lost. Only the \emph{space-time}
intervals between two events have an absolute meaning.

\section{General relativity: Elasticity of time}
\label{s3}

In 1907, while writing a review article on special relativity,
Einstein realized that while his 1905 space-time model successfully
reconciled the predictions of Maxwell's electrodynamics, it was in
deep conflict with Newton's theory of gravity. For, Newton's law of
universal gravitational attraction required absolute time: the
gravitational force between two bodies is inversely proportional to
the square of the distance between them, measured at an instant of
this absolute time. As the earth moves around the sun, the Newtonian
gravitational force it exerts on the moon changes instantaneously.
Special relativity, on the other hand, had abolished absolute time.
So Newton's law can not even be stated in Einstein's 1905 framework.
In 1907, then, Einstein set himself the task of finding a new,
better theory of gravity. After eight years of concentrated effort,
he finally completed it in November 1915. The notions of space and
time changed again, and even more dramatically than ever before.

In general relativity, space and time continue to form a
4-dimensional continuum. But now the geometry of this continuum is
{\it curved} and the amount of curvature in a region encodes the
strength of gravity there. Space-time is not an inert stage on which
things happen. Rather, space-time geometry acts on matter and can be
acted upon. There are no longer any spectators in the cosmic dance.
The stage itself joins the troupe of actors. This is a profound
paradigm shift.

Einstein was motivated in this work by two seemingly simple
observations. First, as Galileo demonstrated through his famous
experiments at the leaning tower of Pisa, the effect of gravity is
universal: all bodies fall the same way if the only force on them is
gravitational. Second, gravity is {\it always} attractive. This is
in striking contrast with, say, the electric force where unlike
charges attract while like charges repel. As a result, while one can
easily create regions in which the electric field vanishes, one can
not build gravity shields. Thus, gravity is omnipresent and
non-discriminating; it is everywhere and acts on everything the same
way. These two facts make gravity unlike any other fundamental force
and suggest that gravity is a manifestation of something deeper and
universal. Since space-time is also omnipresent and the same for all
physical systems, Einstein was led to regard gravity not as a force
but a manifestation of space-time geometry. Space-time of special
relativity is rigid, flat, and fixed once and for all. Space-time of
general relativity is supple and can be `bent' by massive bodies.
The sun for example, being heavy, bends space-time significantly.
Planets like earth move in this curved geometry. In a precise
mathematical sense, they follow the simplest trajectories called
geodesics
---generalizations of straight lines of the flat geometry of Euclid
to the curved geometry of Riemann. So, when viewed from the
\emph{curved space-time} perspective, earth takes the straightest
possible path. But since space-time itself is curved, the trajectory
appears elliptical from the \emph{flat space and
absolute time} perspective of Euclid and Newton.%
\footnote{In general relativity, space-time is again represented by
a 4-dimensional continuum but it need not be topologically
$\mathbb{R}^4$. In cosmology, the actual topology of the entire
universe is directly correlated to the total matter density. Because
the metric is allowed to be curved, subject only to Einstein's
equations, entirely new possibilities can open up and some of them
are known to be realized in Nature: the universe can be expanding
(or contracting); black holes can form through a dynamical collapse
of matter under gravitational attraction; and ripples in the
geometry of space-time can propagate as gravitational waves,
carrying physical attributes such as energy, momentum and angular
momentum! Since the focus of this article is on time, for simplicity
I have restricted myself to globally hyperbolic space-times so that
the initial value problem of dynamics is well posed.}

As a consequence of this paradigm shift, time in general relativity
is less rigid, more `elastic', than that defined even in special
relativity. In special relativity each inertial observer assigns a
well-defined time interval between any two space-time events. In
general relativity, because the space-time geometry is curved and
therefore inhomogeneous, we no longer have the three-parameter
family of global inertial observers. What about the absolute
\emph{space-time} intervals between distant events of special
relativity? In general relativity they are no longer absolute but
depend on the \emph{path joining them.} These contrasts can be
summarized (in a semi-heuristic fashion) as follows. Newton
postulated that time flows in an absolute manner, indifferent to the
observers carrying out measurements or to the physical content of
the universe. Special relativity taught us that time intervals
between distant events depend on the state of motion of the
observers who measure them. General relativity taught us that they
depend, in addition, on the location of the observer. Gravity curves
geometry, making clocks tick at different rates at different
locations: All accurate clocks tick a little faster at the summit of
Mont Blanc than they do in Paris. And these are \emph{real, physical
effects}. The Global Positioning Satellites (GPS) system that we all
use, for example, has to take into account the special relativistic
effects because the clocks on various satellites are moving relative
to us, on earth, and relative to each other. They have to take into
account the general relativistic effect because space-time is bent a
little less at the location of the satellite than it is on earth.
(For details, see, e.g., \cite{na}.) If these effects were ignored,
the GPS system would fail in its navigational functions within about
2 minutes!

To summarize, with each generalization the notion of time has become
less rigid, less absolute, more elastic, more fluid. Time continues
to retain its physical reality that Newton first attributed to it.
However, contrary to the mental image that most people have,
physical time does not march steadily, uniformly indifferent to
everything else.

General relativity also has a profound implication for the global
nature of time. In Newtonian mechanics and special relativity, time
runs from eternal past to the eternal future. In general relativity,
space-time geometry is dynamical. And the `singularity theorems' of
general relativity inform us that so long as matter satisfies local
positive energy conditions, the evolving geometry would develop
singularities in a finite amount of time  in generic cosmological
situations (see, e.g., \cite{he}). In particular, if we impose the
requirement of large scale homogeneity and isotropy of space
---which has been observed to an excellent degree of accuracy---
then Einstein's equations imply that our expanding universe must
have started from a singularity ---the Big Bang--- at a
\emph{finite} time in the past. Within general relativity, it is
meaningless to ask for what was there before. It is \emph{not} that
time was ticking away steadily and matter came into existence with a
Big Bang at this instant. Rather, space-time geometry itself was
born at the Big Bang together with matter. And, since it is this
geometry that enables us to speak of time intervals, \emph{within
general relativity} it is simply meaningless to ask what was there
before the Big Bang.

\section{Beyond general relativity?}
\label{s4}

However, it is commonly accepted that general relativity is
incomplete because it ignores quantum physics. In solutions to
Einstein's equations, if we run the movie of the history of the
universe backward in time and approach the Big Bang, we find that
the density of matter increases continuously and we reach the Planck
regime in which the characteristic scales are set by the
\emph{fundamental} constants of Nature, $G, \hbar$ and $c$. To
arrive at the Big Bang we have to keep going backward in time,
further increasing matter density and curvature until they become
infinite. Now, already in the Planck regime the matter density is
approximately $10^{92}$gm/cc, which is some eighty(!) orders of
magnitude larger than the nuclear density encountered, for example,
in neutron stars. Even at the nuclear density quantum mechanics is
absolutely vital e.g., for the very existence of neutron stars.
Therefore, it is completely inappropriate to ignore the quantum
properties of matter in the Planck regime. Quantum matter is
described by quantum fields. Since matter appears on the right hand
side of Einstein's equations, the description of the left hand side
representing geometry should also incorporate the quantum paradigm.
So, the use of classical Einstein's equations in (and beyond) the
Planck regime is an approximation that has no justification at all.
Consequently, \emph{singularities such as the Big Bang are
predictions of general relativity in a domain in which
it is simply invalid.}%
\footnote{Einstein was well aware of this. On the issue of whether
the prediction of the Big Bang singularity of general relativity is
trustworthy, he wrote \cite{ae}: ``One may not assume the validity
of field equations at very high density of field and matter and one
may not conclude that the beginning of the expansion should be a
singularity in the mathematical sense."}
In spite of the standard rhetoric, it is not at all clear that our
universe really originated in the Big Bang or had any singularity in
the past. Thus, while general relativity does predict that our
universe had a finite beginning, to take that prediction seriously
would be as grossly inadequate as using Newtonian mechanics in the
CERN accelerator experiments, or, using the flat space-time of
special relativity to describe black holes!

To find what \emph{really} happened in the Planck era of the very
early universe, one needs a unification of general relativity and
quantum physics, at least in the setting of cosmology. Attempts at
constructing quantum cosmology theories date back at least to the
early 1970s, when Misner, DeWitt and Wheeler introduced influential
ideas to probe the effects of quantum physics in the very early
universe \cite{cwm,jaw}. By now there are detailed calculations,
particularly in loop quantum cosmology (LQC) \cite{mb,as}, that
strongly suggest that the Big Bang singularity \emph{is} resolved
once general relativity is replaced by a more fundamental theory
based on \emph{quantum} Riemannian geometry.

What about the issue of the Beginning, then? As has happened
repeatedly in the past ---e.g. when Newtonian mechanics was replaced
by special relativity, or the flat, inert space time of special
relativity was replaced by the curved, dynamical space-time of
general relativity--- the new paradigm brings out the necessity of
sharpening the questions one asks, and the answers can vary greatly
depending on which precise version of the question one chooses. As
for the issue of the Beginning, we have at least two possibilities.
One can inquire if, as we evolve backward in time, the equations
break down and the evolution can no longer be continued. If this
occurs,
one can say that space-time and matter originated at that event;
this is the Beginning. In general relativity, of course, this event
is the Big Bang. In LQC, on the other hand, the \emph{quantum}
evolution never breaks down. The quantum state never encounters
infinities; none of the physical observables ever become singular
\cite{as}. In this sense, then, time does \emph{not} have a finite
beginning in LQC. But, alternatively, one could also ask whether
there is an epoch in which the quantum fluctuations in geometry
cannot be ignored.%
\footnote{In loop quantum cosmology, there is such an epoch in which
there are large deviations from general relativity leading to
potentially observable effects \cite{ps}.}
If there is, then the space-time continuum obeying Einstein's
equations would become a trustable approximation only at the end of
that epoch. Then one would say that the notion of time a la general
relativity  `emerged' at the end of that epoch; in this sense, time
has a finite beginning. In LQC there is such an epoch. Therefore, in
this second sense, time did have a finite beginning in LQC.

So far, in this section  I have focused only on the global question
of whether time is likely to have a finite beginning in quantum
gravity theories that go beyond general relativity. What about the
more local issue of time and space-time intervals, discussed in the
last three sections? In quantum gravity theories, smooth space-times
of general relativity would not be fundamental. The fundamental
objects of the theory ---`degrees of freedom' in the physics
literature--- may be quite different from the space-time metric of
general relativity. In loop quantum gravity, for example, the
fundamental excitations of gravity/geometry are polymer-like with
one space and one time dimension, encoded in `spin networks' and
`spin foams' \cite{alrev,crbook,ttbook}. The space-time continuum
would emerge, perhaps as a condensate \cite{gft} of these more
fundamental quantum building blocks, if one coarse grains, i.e.,
ignores physics below, say, $10^{-15}$cm, the length scale used in
particle physics. Therefore the space-time intervals between two
events along specified curves, used in general relativity, are also
approximate concepts. But in general relativity, we need the notion
of such intervals to speak of dynamics and make predictions. Without
access to these notions, how would one discuss dynamics in quantum
gravity? This is a difficult question because, to face it
adequately, one must treat the clocks themselves using quantum
mechanics \cite{gp}. The `issue of time in quantum gravity' is still
very far from being fully understood and therefore a subject of
active discussion (for a summary, see, e.g., \cite{as-book}).

I will conclude by summarizing what has proved to be the most
successful strategy so far. Interestingly, it calls on us to return
to the notion of relational time I began with. The primary emphasis
is on \emph{interdependence}, advocated by Ernst Mach. The
idea is to treat one of the dynamical variables ---e.g., a scalar
field, or the total matter density, or a suitable curvature scalar,
or, ...--- as the `clock variable' with respect to which all other
physical observables change. Thus, the questions one poses are of
the following type. Suppose that the quantum state is
such that, when the scalar field $\phi$ (treated as the clock
variable) has the value $\phi_1$, the expectation value and quantum
uncertainties of observables of interest ---e.g., the scalar
curvature and anisotropy parameters--- have such and such values.
What would their values be when the scalar field $\phi$ takes the
value $\phi_2$? This change is interpreted as `evolution' of the
chosen observable (scalar curvature or anisotropy variables) with
respect to the chosen relational time variable (the scalar field).
For this qualitative idea to work in quantitative detail, the clock
variable has to be chosen astutely. In simple cosmological models,
there are natural choices, the strategy has been implemented in
detail, and, as one would hope, the results reduce to those from
general relativity once one is well outside the Planck regime.
However, one faces the following obvious issue: How do you compare
predictions from two distinct choices of the clock variables in the
full quantum gravity regime?

To probe this question, it is useful to begin by discussing the
issue of time within general relativity in greater detail than I
have done so far. Recall first that, although we do not have an
absolute time in special relativity, we can associate a notion of
time with each inertial observer; this is only a 3 parameter
freedom. We describe dynamics using the time variable adapted to an
inertial observer and at the end verify that the physical results
are insensitive to the choice. In general relativity we do not have
global inertial observers. Therefore the ambiguity becomes infinite
dimensional. To discuss dynamics, one first slices space-time into
space and time using \emph{any} family of space-like (Cauchy)
surfaces. Each of these slices now represents a possible `instant of
time' and time evolution now refers to the change of physical fields
with respect to the coordinate labeling these slices. Thus, given
initial data for, say, the electromagnetic field at an initial time,
i.e., on one of these space-like slices, one can evolve them using
Maxwell's equations on the given curved space-time and obtain a
solution on that entire space-time. This procedure makes heavy use
of the underlying space-time metric. But the setup guarantees that
the final solution is independent of the choice of slicing in the
following precise sense. Suppose we have two slicings, leading to
two time variables $t$ and $t'$. For simplicity let us suppose that
they agree for, say $t < t_1$ and $t> t_2$ for some $t_1, t_2$ with
$t_2 > t_1$, but differ over the intermediate times between $t_1$
and $t_2$. Then, the two evolutions carried out using $t$ and $t'$
from any given initial data on any slice $t = {\rm const} <t_1$ will
lead to the \emph{same} initial data on any slice with $t = {\rm
const}> t_2$. Furthermore, even in the intermediate region where the
two slices are distinct, if we were to evaluate the initial data of
the solution obtained using the $t$ evolution on any $t'= {\rm
const}$ slice, we would find a complete agreement with the initial
data on that slice provided by the  $t'$ evolution. Thus, although
the notion of time evolution in general relativity is nowhere as
unique or canonical as that in Newtonian mechanics or special
relativity because time in general relativity
is\emph{`multi-fingered'}, the coherence we just discussed ensures
that all permissible evolutions are physically equivalent. In the
language of the barter system I began this article with, the
coherence guarantees that if I started out with $n$ sheep and traded
them for chicken and then traded chicken for oil and then traded oil
back for sheep, (in an ideal world without profits) I would again
obtain the same number $n$ of sheep I started with. There is a
global consistency to the scheme. In this commercial example, it is
this coherence that enables one to introduce an abstract notion of
money and supplant the barter system by assigning a well-defined
monetary value to each of our goods. Returning to general
relativity, the coherence lets one attribute a physical reality to
the notion of time and evolution although, being `multi-fingered'
the notion is now much more sophisticated than that in special
relativity.

With relational observables in quantum gravity, the situation is not
so simple. First, we no longer have a space-time continuum that can
be foliated by space-like slices because there are regimes in which
there would be no coarse graining whatsoever that would lead to a
smooth space-time metric. Second, each choice of a clock variable
naturally leads to a set of observables whose evolution can be
studied in a natural manner (e.g., without having to make ad-hoc
factor ordering choices for operators), and the set can vary from
one choice to another. A related but distinct problem is that each
procedure endows the space of solutions to the quantum Einstein
equations with a specific and distinct Hilbert space structure with
which one computes expectation values and fluctuations of
observables. As a result, in general it is quite difficult to
compare dynamical predictions in quantum theories arising from two
different choices. In the simplest cases where this is feasible, one
does find consistency \cite{atu}. However, the general problem is
completely open. As a result, it is not clear whether, when the dust
settles, there will be the appropriate `coherence' between different
choices to say that there is an underlying, objective (but vastly
generalized) notion of time that the quantum dynamics refer to. It
may happen that, for each (admissible) choice of relational time, we
have well-defined and internally consistent dynamics. But rather
than fusing together to give us a single, globally coherent
paradigm, these theories could remain distinct. Furthermore, it may
be that each of them passes its own observational tests. This could
happen because the observables measured to test one of these
theories fail to be well-defined in another! In that case, time
would be relational in its core and the primary focus would be on
correlations between physical quantities. Leibnitz's `relational' notions and Mach's `interdependence' would then constitute the very essence of fundamental physics.

\section*{Acknowledgments} This article is an extended version of
my contributions to the discussion sessions of the Penn State
Workshop on Time and Cosmology and owes its existence to the persuasive
skills of Emily Grosholz. I would like to thank Bryan Roberts and Tom
Pashby for comments on the first draft. This work was supported in part by
the NSF grant PHY-1205388 and the Eberly research funds of Penn State.
Parts of the historical discussion in sections \ref{s2} and
\ref{s3} overlap with another of my articles \cite{aa}.%
\footnote{That article appeared in a collection which,
unfortunately, is not easily accessible. Moreover, the primary focus
of that article was on the issue of singularities, rather than
time.}

\end{document}